\documentclass[aps,twocolumn]{revtex4}
\usepackage{enumerate}
\usepackage{amsfonts}
\usepackage{graphicx}
\usepackage{dcolumn}
\usepackage{bm}
\usepackage{amsmath}
\usepackage{amsxtra}
\usepackage{amstext}
\usepackage{amssymb}
\usepackage{latexsym}

\begin{document}
\title{Family of Probability Distributions Derived from Maximal Entropy Principle with Scale Invariant Restrictions}

\author{Giorgio Sonnino$^{1*}$, Gy\"{o}rgy Steinbrecher$^2$, Alessandro Cardinali$^3$, Alberto Sonnino$^4$, Mustapha Tlidi$^1$}
\address{$^1$ Department of Theoretical Physics and Mathematics, Universit{\'e} Libre de Bruxelles (U.L.B.), 
Campus de la Plaine C. P. 231 - Boulevard du Triomphe, 1050 Brussels, Belgium}
\address{$^2$ Association EURATOM-MEC, Physics Faculty, University of Craiova, Str.A.I.Cuza 13, 200585 Craiova, Romania}
\address{$^3$ EURATOM-ENEA Fusion Association, Via E.Fermi 45, C.P. 65 - 00044 Frascati (Rome), Italy}
\address{$^4$ Universit{\'e} Catholique de Louvain (UCL), Ecole Polytechnique de Louvain (EPL), Rue Archim$\grave{\rm e}$de, 1 bte L6.11.01, 1348 Louvain-la-Neuve, Belgium}
\begin{abstract}
Using statistical thermodynamics, we derive a general expression of the stationary probability distribution for thermodynamic systems driven out of equilibrium by several thermodynamic forces. The local equilibrium is defined by imposing the minimum entropy production and the maximum entropy principle under the scale invariance restrictions. The obtained probability distribution presents a singularity that has immediate physical interpretation in terms of the intermittency models. The derived reference probability distribution function is interpreted as time and ensemble average of the real physical one. A generic family of stochastic processes describing noise-driven intermittency, where the stationary density distribution coincides exactly with the one resulted from entropy maximization, is presented. 

\vskip0.2truecm


\vskip0.3truecm
\noindent *Email: gsonnino@ulb.ac.be
\end{abstract}

\maketitle

\noindent In the Onsager region, the minimum entropy production (MEP) theorem provides a variational characterization of a stationary states, both for macroscopic systems and for stochastic models. However, to characterize some unknown events with a statistical model, we should choose the one that has maximum entropy (MaxEnt's principle). This means that, out of all probability distributions consistent with a given set of constraints, we choose the one that has maximum uncertainty. Entropy is therefore regarded as a measure of information. The MaxEnt principle developed in communication and in information technology has recently been found to have a wide ranging applications in many areas of science. In evolutionary biology, the MaxEnt principle is used to explain the observed species abundance distribution. In this case the constraint may be set by the habitat, which fixes the average population size of the species \cite{shipley}-\cite{frank}. This principle is also applied for describing the activity in a variety of neural networks \cite{shlens}-\cite{ganmor}, the statistics of amino acid substitutions in protein families \cite{seno}-\cite{sulkowska} and in material science \cite{nakazawa} and in Bose-Einstein condensate in a dye microcavity \cite{Denis-sob}. It is also proven that the MaxEnt model can provide a good description of data from seemingly completely unrelated phenomena observed in social contexts \cite {baek}, \cite{hernando}. For example, it describes very well the equilibrium distribution function for the city-population subject to two additional scale-invariant constraints: the normalization of the probability distribution function and the expected value of the city-population \cite{hernando1}. 

\noindent The purpose of this paper is to use statistical thermodynamics to derive a reference density distribution function (DDF) for open thermodynamic systems close to a local equilibrium state. To this end, we assume the validity of the minimum entropy production principle and the MaxEnt principle on the random variables, under the scale invariance restrictions. By definition, the reference DDF (indicated with ${\mathcal F}^0$) is an initial distribution function. ${\mathcal F}^0$ should depend only on the invariants of motion, with the property to evolve slowly from the local equilibrium state i.e., it remains confined for sufficiently long time. Hence, the reference DDF results in a perturbation of the local equilibrium state. We consider open thermodynamic systems obeying to Prigogine's statistical thermodynamics. We then define the local equilibrium state by adopting a minimal number of hypotheses. Finally, we link the density distribution function with particle$'$s DDF. The density probability distribution of finding a state in which the values of the fluctuating thermodynamic variable, ${\tilde\beta}_\kappa$, lies between ${\tilde\beta}_\kappa$ and  ${\tilde\beta}_\kappa+d{\tilde\beta}_\kappa$ is
\begin{equation}\label{i1}
{\mathcal F}={\mathcal N}_0\exp[-\Delta_I S]
\end{equation}
\noindent where ${\mathcal N}_0$ ensures normalization to unity, and we have introduced the dimensionless (density of) entropy production $\Delta_I S$. The negative sign in Eq.~(\ref{i1}) is due to the fact that, during the processes, $-\Delta_IS\leq0$. Indeed, if $-\Delta_IS$ were positive, the transformation ${\tilde\beta}_\kappa\rightarrow {\tilde\beta}'_\kappa$ would be a spontaneous irreversible change and thus be incompatible with the assumption that the initial state is a {\it stable} (local) equilibrium state \cite{prigogine}. We suppose that the system is subject to ${\tilde N}$ thermodynamic forces. The entropy production $\Delta_IS$ is linked to the thermodynamic forces $X^\kappa$, the thermodynamic fluctuation ${\tilde\beta}_\kappa$, and the thermodynamic flows $J_\kappa$ by the following set of equations \cite{degroot}
\begin{equation}\label{i2}
X^\kappa=\frac{\partial \Delta_IS}{\partial{\tilde\beta}_\kappa}\ \ , \ \ J_\kappa=\frac{d{\tilde\beta}_\kappa}{dt}\ \ , \ \frac{d_IS}{dt}=\sum_{\kappa=1}^{\tilde N}  X^\kappa J_k\!\geq 0\\
\end{equation}
\noindent Our aim is firstly to derive the reference DDF, ${\mathcal F}^0$, by imposing the MEP and the MaxEnt principle, submitted to the scale invariance constraints. Successively we shall determine the class of stochastic processes whose stationary PDF includes ${\mathcal F}^0$ as a special case. 


\noindent Let us consider an open thermodynamic system subject to ${\tilde N}=N+1$ thermodynamic forces. $N$ thermodynamic forces are linked to $N$ fluctuations of Prigogine$'$s type (i.e., the entropy production is expressed in quadratic form with respect to these fluctuations. For an exact definition of Prigogine$'$s fluctuations refer to \cite{prigogine}, \cite{prigogine1}. One thermodynamic force is linked to a fluctuation of a different nature. This random variable has the density distribution function, which satisfies the MaxEnt principle under the scale invariant restrictions. This fluctuation will be indicated with $w$ whereas the Prigogine fluctuations will be denoted by $\beta_\kappa$, with $\kappa=1, \cdots,, N$. Hence, ${\tilde\beta}_\kappa=\beta_k$ for $k=1,\cdots , N$ and ${\tilde\beta}_{N+1}=w$. To be more precise, we identify the local equilibrium state by imposing the following two conditions.
\noindent 
\begin{itemize}
\item[\bf i)] {\it The local equilibrium state corresponds to the values of the fluctuations $\beta_\kappa$ with $\kappa=1,\cdots, N$ for which the entropy production tends to reach an extreme}.
\end{itemize}
\noindent Under this assumption, close to the local equilibrium, the entropy production can be brought into the form
\begin{align}\label{a1}
&\!\!\!\!\!\!-\Delta_IS=g_0(w)-\frac{1}{2}\sum_{\kappa ,j=1}^Ng_{\kappa j}(w)\beta_\kappa\beta_j+h.o.t.\\
&{\rm with}\qquad g_0(w)\equiv -\Delta_IS\mid_{\beta_1\cdots\beta_N=0}\nonumber
\end{align}
\noindent where $h.o.t.$ stands for {\it higher order terms}. Coefficients $g_{\kappa j}$ are directly linked to the transport coefficients of the system \cite{sonnino}. Therefore, the general expression for ${\mathcal F}$, given by Eq.~(\ref{i1}), becomes the reference DDF, ${\mathcal F}^{0}$, when the entropy production is provided by Eq.~(\ref{a1}). The DDF
\begin{equation}\label{a2}
\mathcal{P}(w)\equiv{\mathcal N}_0\exp[g_0(w)]={\mathcal N}_0\exp[-\Delta_IS\mid_{\beta_1\cdots \beta_N=0}]
\end{equation}
\noindent related to the variable $w$, at $\beta_\kappa=0$ (with $\kappa=1\cdots N$), is determined by the following condition.
\begin{itemize}
\item[\bf ii)] {\it At the extremizing values $\beta_\kappa=0$ with $\kappa=1,\cdots, N$ under the scale invariance restrictions, the system tends to evolve towards the maximal entropy configurations}.
\end{itemize}

\noindent Notice that in particular, if the imposed the two scale invariant restrictions, ${\rm E}[w]=const. >0$ and ${\rm E}[\ln (w)]=const.$ (where ${\rm E}[\ ]$ is the expectation operation), together with normalization, the density distribution for the $w$ variable is given by a {\it gamma distribution function} \cite{papoulis}
\begin{equation}
\mathcal{P}(w)=\mathcal{N}_{0}w^{\gamma-1}\exp(-w/\Theta)\label{0.1}
\end{equation}
\noindent Here, we have introduced the {\it scale parameter} $\Theta$ and the {\it shape parameter} $\gamma$, which could be determined by the fit of experimental data. The goal is now to justify {\it a posteriori} the result found in Ref.~\cite{sonnino} :  {\it The kinetic energy dependence of the DDF}, $\mathcal{P}(w)$, {\it at the most probable (equilibrium) values of the remaining phase space variables, is expressed by the gamma distribution}. Notice that if we normalize $\mathcal{P}(w)$ to unity then it can be interpreted as the conditional probability density function (PDF), $\rho(w)$, conditioned by $\beta_\kappa=0$ with $\kappa=1,\cdots N$ i.e., $\rho(w)\equiv\mathcal{P}(w\mid \beta_1=\cdots=\beta_N=0)$. 

\noindent We consider the entropy $S[\rho(.)]$ of probability density function (PDF), $\rho(w)\geq0$, given by
\begin{equation}
S[\rho(.)]=-\int_{0}^{\infty}\rho(w)\log(\rho(w))dw \label{SG1}
\end{equation}
\noindent We investigate the consequences of the MaxEnt principle submitted to the most general scale-invariant restrictions. Let us then start to consider the following restrictions
\begin{equation}\label{1.1}
\int_{0}^{\infty}w^{\alpha_{k}}\rho(w)dw  =\mathrm{E}(w^{\alpha_{k}}
)=\mu_{k};~k=0,1,\ldots,n
\end{equation}
\noindent where, at this stage, $\alpha_{k},$ and $\mu_{k}$ are real numbers. However, for completeness, we should also include the following restriction obtained as the limit case of (\ref{1.1})
\begin{equation}\label{1.2}
\mathrm{E}(\log(w)) =\nu
\end{equation}
\noindent where $\nu$ is a real number. Indeed, suppose that we have for some fixed $k$ : $\alpha_{k}=\varepsilon\ll1$, then
\begin{equation}
\mathrm{E}(w^{\varepsilon})=\mu_{k}~ \label{SG7}
\end{equation}
\noindent From Eqs~(\ref{1.1}) and by taking into account the normalization condition on the probability [see the forthcoming Eq.~(\ref{SG2})], we get
\begin{equation}
\mathrm{E}\left(  \frac{w^{\varepsilon}-w^{0}}{\varepsilon}\right)  =\frac
{\mu_{k}-1}{\varepsilon} \label{SG8}
\end{equation}
\noindent If the support of the PDF $\rho(w)$ is concentrated mainly on the domain where $|\log(w)|$ is not too large, then we can approximate: $\left(
w^{\varepsilon}-w^{0}\right)  /\varepsilon\cong\log(w)$. In this case Eq.~(\ref{SG8}) reduces to Eq.(\ref{1.2}). Finally, Eq.(\ref{1.2}) can be seen as the {\it limit case} of (\ref{1.1}) and, for the sake of generality, it should be taken into account among the equations expressing the scale-invariance restrictions. It is worthwhile mentioning that the equilibrium distribution function, obtained by imposing restriction (\ref{1.2}), retrieves the expression currently used for fitting the numerical steady-state solution of the simulations for the Ion Cyclotron Radiation Heating (ICRH) FAST-plasmas, and for describing various scenarios of tokamak-plasmas \cite{sonnino}. Another example showing the usefulness of constraint (\ref{1.2}) can be found at the end of this report. Let us now consider the following simplest physical case
\begin{align}
\ \int_{0}^{\infty}\rho(w)dw  &  =1\label{SG2}\\
\ \int_{0}^{\infty}w\rho(w)dw  &  =\mathrm{E}(w)=\mu_{1} \label{SG3}
\end{align}
\noindent Hence, $n\geq1$ and, in particular
\begin{equation}
\mu_{0}\equiv1;~\alpha_{0}=0;\alpha_{1}=1 \label{3.1}
\end{equation}
\noindent where $\mu_{1}=T/m$, with $T$ and $m$ are temperature and particle's mass, respectively. 

\noindent To understand the meaning of Eqs~(\ref{1.1}) and (\ref{1.2}), for $n=1$, we find Eq.~(\ref{0.1}) from the MaxEnt principle submitted to the constraints (\ref{1.2}), (\ref{SG2}) and (\ref{SG3}). In the following, we shall be able to give only a partial answer to this question. The restrictions on the PDF given by Eq.~(\ref{1.1}) and Eq.~(\ref{1.2}) are invariant under scale transformations. This means that  the effect of the scale transformation $w\rightarrow kw^{\prime}$ on the restrictions (\ref{1.1}) and (\ref{1.2}) is $\mathrm{E}(w^{\prime\alpha_{k}})=\mu_{\alpha_{k}}=\mu_{\alpha_{k}}k^{-\alpha_{k}}$ and $\mathrm{E}(\log(w^{\prime}))=\nu^{\prime}=\nu-\log(k)$, respectively. So restrictions (\ref{1.1}) and (\ref{1.2}) remain in the same class specified by the functions $w^{\alpha_{k}}$ and $\log(w)$, respectively.

\noindent {\bf Remark (1)}. \emph{It remains to explain the ansatz}: $n=1$\emph{, and} Eq.~(\ref{1.2}) [\emph{naturally combined with restrictions} (\ref{SG2}) \emph{and} (\ref{SG3})].

\noindent By denoting with $\lambda_{k},0\leq k\leq n+1$ the Lagrange multipliers in the problem of maximizing the entropy given by Eq.~(\ref{SG1}), with the restrictions (\ref{1.1}) and (\ref{1.2}), we get
\begin{equation}
\log\left[  \rho(w)\right]  \ =1-
{\displaystyle\sum_{k=0}^{n}}
\lambda_{k}w^{\alpha_{k}}-\lambda_{n+1}\log(w) \label{3.5}
\end{equation}
\noindent It is convenient to introduce the following parametrization of the general PDF:
\begin{equation}
\rho(w)\ =\frac{1}{Z}w^{\gamma-1}\exp\left[  -
{\displaystyle\sum_{k=1}^{n}}
\lambda_{k}w^{\alpha_{k}}\right]  \ \label{3.6}
\end{equation}
\noindent with $\gamma-1=-\lambda_{n+1}$ and $1/Z=\exp(1-\lambda_{0})$. The parameters $Z$, $\gamma,$ $\lambda_{k}$ are fixed by Eqs~(\ref{1.1}) and (\ref{1.2}). We should have $\lambda_{k}>0$ when $\alpha_{k}$ is the largest exponent, as well as when $\alpha_{k}$ is the least negative exponent.


\noindent In Eq.(\ref{3.6}), despite the condition $\alpha_{k}<0$, with $\lambda_{k}>0$, is mathematically acceptable, we should exclude this possibility from the physical point of view. Because in several situations, like with collisional plasmas, such a condition would provide a very reduced population at low energy. Similarly, also the alternative $\alpha_{k}>1$, with $\lambda_{k}>0$, for some value of $k$, should be discarded. This because such a condition would give rise to a {\it sub-Maxwellian} distribution function when, on the contrary, it is expected that the high energy tail of the PDF should be {\it larger} than the one predicted by the Maxwellian distribution (consider, for instance, the case of burning plasmas).

\noindent {\bf Conjecture (2)}. \emph{Physical considerations lead us to select} $\max(\alpha_1,\cdots,\alpha_n=1$, and $0), \alpha_k<1$ if $k>2$.


\noindent Even though we accept the condition $\alpha_{k} <0$, the case $\alpha
_{k}\rightarrow0$, should anyhow be considered as a limit situation because the graph of
$\rho(w)$ changes abruptly with the change of the sign of $\alpha_{k}$. From this reason, the restriction (\ref{1.2}) can be considered as
an extreme case. Consequently the partial justification of the ansatz, in the Remark (1), is given for simplicity and extremality reasons. The
exploration of the case $n=2$, with $0<\alpha_{2}<1$ is ongoing. It is
clear that in this simplest and extremal case, given by the ansatz in the Remark (1), the PDF has the form
\begin{equation}
\rho(w)=\frac{1}{\Theta\Gamma(\gamma)}\left(  w/\Theta\right)  ^{\gamma-1}
\exp\left(  -w/\Theta\right)  \label{SG9}
\end{equation}
\noindent where $\Gamma(\gamma)$ is the Euler Gamma function. The relations between the parameters $\gamma$, $\Theta$, $\mu_{1}=T/m$ and $\nu$ are given by
\begin{align}
\ \ \mathrm{E}(w)  &  =T/m=\gamma\Theta\label{SG10}\\
\ \mathrm{E}(\log(w))  &  =\nu=\Psi(\gamma)+\log(\Theta) \label{SG11}
\end{align}
\noindent where $\Psi(\gamma)$ is the digamma function. We draw the attention to the fact that, in general, the parameters appearing in the Eqs~(\ref{1.1}) and (\ref{1.2}) are not completely independent. Indeed suppose that we have
\ $\mathrm{E}(w^{\alpha_{1}})=\mu_{1}$ and $\mathrm{E}(w^{\alpha_{2}})=\mu_{2}$ with $\alpha_{1}<\alpha_{2}$. Then, by using the H\"{o}lder inequality
\cite{Rudin}, we obtain:
\begin{equation}
\mu_{1}^{1/\alpha_{1}}\leq\mu_{2}^{1/\alpha_{2}}\label{SG14}
\end{equation}
\noindent In the limit case, by using Eqs~(\ref{1.2}) and (\ref{SG3}), together with Jensen's inequality for the exponential function \cite{Rudin}, we find
\begin{equation}
\mu_{k}^{1/\alpha_{k}}\geq\exp\left(  \mathrm{E}(\log(w))\right)
=\exp(v);k\geq1\label{SG15}
\end{equation}
\noindent The inequalities (\ref{SG14}) and (\ref{SG15}) are the {\it necessary conditions
for the existence} and {\it not} the necessary conditions for maximal entropy of the PDF.  The inequality (\ref{SG14}) results from the constraint (\ref{1.1}) whereas (\ref{SG15}) from the constraints (\ref{1.1}) and (\ref{1.2}). In
particular, for restrictions (\ref{SG2}) and (\ref{SG3}) we get $\mu_{1}\geq\exp\left(  v\right)$.

\noindent Due to the very special choice of the restrictions used in the MaxEnt principle (the minimality of the $\log(w)$), the
resulting PDF is expected to have some special properties. Indeed, the gamma distribution is \emph{infinitely divisible} and \emph{stable} in the following sense. Let $X_{1},\ldots,X_{n}$ be a sequence of $n$ independent random variable having $n$ identical gamma
distributions (with the same scale parameter and shape parameters $\gamma
_{1},\ldots\gamma_{n}$), then the sum $\sum_{k=1}^{n}X_{k}$
has the same gamma distribution function, with the same scale parameter and with shape
parameter equal to $\sum_{k=1}^{n}\gamma_{k}$. An important physical property of the distributions (\ref{3.6}), in the case $0<\gamma<1$, is the following. By varying $\gamma$ in this range of parameters, and by keeping constant the other parameters, the shape of the graph of the DDF changes, due to the singularity appearing at $w=0$. This singularity can be associated with the appearance of a regime of intermittent behavior \cite{Aumaitre}. We emphasize that the reference DDF is interpreted as time and ensemble average of the physical, DDF. 


\noindent We describe a class of stochastic processes admitting Eq.~(\ref{3.6}) as stationary PDF solutions. Note that Eq.~(\ref{3.6}) reduces to the gamma distribution when some coefficients are set to zero. To this end, we consider the stochastic differential equation (SDE) for the random variable $w(t)$. In this case $w(t)$ is the energy of an individual charged particle. The SDE includes the simplest soluble cases of the class of intermittency models \cite{Schenzle}-\cite{Aumaitre2}. In the Stratonovich version, the SDE reads 
\begin{equation}
dw(t)=(adt+\sigma dB(t))\circ w(t)-S\left[  w\left(  t\right)  \right]dt\label{z1}
\end{equation}
\noindent where $\circ$ stands for the {\it Stratonovich product}. In addition, $a>0$ is the instability threshold, $B(t)$ is the standard Brownian motion (Wiener's process) and  $\sigma$ is the intensity of the multiplicative noise. The function $S(w)$ is associated with the saturation of the instability controlled by the linear term. Hence, we should have
\begin{equation}
\underset{w\rightarrow\infty}{\lim}\frac{S(w)}{w}=+\infty\label{z1.1}
\end{equation}
\noindent We also require that the solution near $w=0$ is dominated by the linear term. So the phenomenology described by Eq.~(\ref{z1}) is still related to the
noise-driven intermittency if we require that
\begin{equation}
\underset{w\rightarrow0+}{\lim}\frac{S(w)}{w}=0\label{z1.2}
\end{equation}
\noindent We will see that, in the particular case $S(w)=Aw^{2}$, the stationary solution of Eq.~(\ref{z1}) is the gamma distribution. Hence, the class of Eqs~(\ref{z1}) includes the generic family of equations describing the instability growth, on the positive semi-axes (which corresponds to our case), limited by the saturation term. We have slightly modified this equation by adding the random multiplicative noise term $\sigma dB(t)$. In the It\^{o} formalism we get
\begin{align}\label{z1a}
&\!\!\!\!\!\!\!\!dw(t) =(a^{\prime}dt+\sigma dB(t))w(t)-S\left[  w\left(  t\right)
\right]dt\nonumber\\
&\!\!\!\!\!\!\!\!{\rm with}\qquad a^{\prime} =a+\frac{\sigma^{2}}{2}
\end{align}
\noindent The stationary Fokker-Planck equation for the density distribution $\rho(w)$ reads
\[
\frac{\partial}{\partial w}\left[  (a^{\prime}w-S(w))\rho(w)\right]
-\frac{\sigma^{2}}{2}\frac{\partial^{2}}{\partial w^{2}}\left[  w^{2}
\rho(w)\right]  =0
\]
\noindent admitting, up to a normalization constant, the following steady state solution
\begin{align}
\rho(w) &  =Cw^{\overline{\gamma}-1}\exp\left[  -\int\frac{S(w)}{w^{2}
}dw\right]  \label{z2}\\
\overline{\gamma} &  =\frac{2a}{\sigma^{2}}>0\label{z2.1}
\end{align}
\noindent The general form of $S(w)$, compatible with Eqs~(\ref{z1.1}) and (\ref{z1.2}), is
\begin{equation}\label{z4}
S(w) =\sum_{k=1}^{m}A_{k}w^{1+\xi_{k}}\qquad {\rm with}\qquad\xi_{k} >0
\end{equation}
\noindent with the constraint that at infinity the coefficient of the leading term in Eq.(\ref{z4}) should be positive. From Eqs~(\ref{z2}) and (\ref{z4}) we obtain
\begin{equation}
\rho(w)=Cw^{\overline{\gamma}-1}\exp\left[-\sum_{k=1}^{m}\frac{A_{k}}
{\xi_{k}}w^{\xi_{k}}\right]  \label{z5}
\end{equation}
\noindent By comparing Eq.~(\ref{z5}) with Eq.~(\ref{3.6}), we can link the exponents
$\alpha_k$ with the exponents $\xi_{k}$ appearing in Eq.(\ref{z4}). Notice that, by setting the parameters entering in the
Eqs~(\ref{z1}) and (\ref{z4}) as 
\begin{equation}\label{z7}
\overline{\gamma}  =\frac{2a}{\sigma^{2}}=\gamma\qquad{\rm and}\qquad \alpha_{k}  =\xi_{k}
\end{equation}
\noindent we obtain, by intermittence mechanism, exactly the same stationary distribution function derived by the MaxEnt principle under the scale invariant restrictions. In addition, we have $\lambda_{k}=A_{k}/\xi_{k}$. Let us now consider the standard Landau type generic form for modeling the instability growth. In the particular case of the simplest choice $S(w)=Aw^{2}$, we obtain the PDF from Eq.(\ref{0.1}) resulting from the Ansatz in the Remark (1). The high energy tail is in agreement with the Maxwell distribution. However, for the case of the Landau term $S(w)=Aw^{3}$, we obtain a
{\it distorted} Gaussian distribution
\begin{equation}
\rho(w)=Cw^{\overline{\gamma}-1}\exp\ \left( -Aw^{2}\right)\label{z8}
\end{equation}
\noindent where the high energy tail decreases {\it too quickly}. This behaviour is incompatible with many physical situations (such as, for example, the case of collisional plasmas) and, for this reason, this possibility should be discarded. Instead, in accordance with Conjecture (2), if in the leading terms $A_\kappa w^{1+\xi_\kappa}$, we have $\xi_1=1$ and for the rest $0<\xi_k<1$, the high energy tail would decrease in a {\it good way}. Hence, Conjecture (2) is supported by the intermittency models. Notice that in the special case $n=0$, in Eqs.~(\ref{3.6}, \ref{z5}) the resulting DDF is non-integrable, but self similar, like in the case of self organized criticality (SOC) models \cite{Turcotte}, \cite{PerBak}. So restriction given by Eq.~(\ref{1.2}) is a remnant SOC, avalanche-like behavior. By replacing the self similarity of the DDF by scale invariance of the constraints in the MaxEnt principle, we obtain a class of DDF that better parametrize the intermittent, avalanche- like behaviour \cite{Turcotte}, \cite{Budaev} of the physical systems that can be modeled by self organized criticality models \cite{Gruzinov}.

\noindent We conclude this brief report by summarizing the main results. Using statistical thermodynamics, we derive the general expression of the reference PDF, ${\mathcal F}_0$, by imposing the minimum entropy production, and the maximum entropy principle under the scale invariance restrictions. Some restrictions on the parameters entering in ${\mathcal F}_0$ are obtained by physical arguments. The obtained PDF exhibits a singularity, which has immediate physical interpretation in terms of the intermittency models. Finally, we derived a family of stochastic processes admitting ${\mathcal F}_0$ as stationary PDF solutions.

\noindent This work gives several perspectives. Through the thermodynamical field theory (TFT) \cite{sonnino1} it is possible to estimate the PDF when the nonlinear contributions cannot be neglected \cite{sonnino2}. The next task should be to establish the relation between the reference PDF herein derived with the one found by the TFT. The solution of this difficult problem will contribute to provide a link between a microscopic description and a macroscopic approach (TFT). Another problem to be solved is the possibility to improve the numerical fit by adding new free parameters according to the principles exposed in this report. 
\vskip0.3truecm
\noindent {\bf Acknowledgements}

\noindent G. Sonnino is very grateful to M.Malek Mansour, of the Universit{\'e} Libre de
Bruxelles, for his scientific suggestions and for his help in the
development of this work. M.Tlidi is a Research Associate with the Fonds de
la Recherche Scientifique F.R.S.-FNRS, Belgium.

\end{document}